\documentclass[runningheads]{llncs}

\usepackage{listings}
\usepackage{xcolor}

\usepackage{graphicx}
\usepackage{algorithm}
\usepackage{algpseudocode}

\usepackage{caption}
\usepackage{subcaption}
\usepackage{enumitem}

\usepackage{pgfplots}
\pgfplotsset{width=10cm,compat=1.15}
\pgfkeys{/pgf/number format/fixed}

\pgfplotsset{
    width=\textwidth,
    height=3cm,
    compat=1.9,
    scaled y ticks=true,
    label style={font=\tiny},
    tick label style={font=\tiny}
}

\usepackage{tabularx} 
\usepackage{booktabs} 

\newcommand{\Name}{ARISE}

\begin{document}
\title{\Name{}: Automating RISC-V Instruction Set Extension\thanks{This work has been founded by the ZuSE project Scale4Edge. Scale4Edge is funded by the German ministry of education and research (BMBF)}}
%
%
\author{Andreas Hager-Clukas\inst{1}\orcidID{0009-0004-9183-6985} \and
Philipp van Kempen\inst{2}\orcidID{0000-0002-1135-8070} \and
Stefan Wallentowitz\inst{1}\orcidID{0000-0003-3182-4929}} 

\authorrunning{A. Hager-Clukas et al.}

%
\institute{Hochschule München University of Applied Sciences, \email{andreas.hager-clukas@hm.edu} \and 
Technical University of Munich}
\maketitle              
%


\begin{abstract}

    RISC-V is an extendable Instruction Set Architecture, growing in popularity for embedded systems. 
    However, optimizing it to specific requirements, imposes a great deal of manual effort.
    To bridge the gap between software and ISA, the tool \Name{} is presented. 
    It automates the generation of RISC-V instructions based on assembly patterns, which are selected by an extendable set of metrics.
    These metrics implement the optimization goals of code size and instruction count reduction, both statically and dynamically.
    The instruction set extensions are generated using the ISA description language CoreDSL.
    Allowing seamless embedding in advanced tools such as the retargeting compiler \textit{Seal5} or the instruction set simulator \textit{ETISS}. 
    \Name{} improves the static code size by 1.48 \% and the dynamic code size by 3.84 \%, as well as the number of instructions to be executed by 7.39 \% on average for \textit{Embench-Iot}.
     
\end{abstract}

\section{Introduction}
\label{sec:Introduction}


In recent years, there has been a growing demand for reducing power consumption in edge devices.
A potential solution to this problem is the reduction of data movement on the core, which can be accomplished by decreasing the size of the code and the number of instructions executed.
Achieving this objective necessitates the implementation of an Instruction Set Architecture (ISA) tailored to the target software.
This customization aims to optimize static and dynamic code size, as well as the dynamic instruction count. 

This paper presents \Name{}, a tool designed for the automated generation of RISC-V Instruction Set Extensions (ISE). 
The objective of the tool is to facilitate and expedite the design and adaptation process for novel ISEs.
A RISC-V specific implementation enables the generation of new instructions, taking into account the architecture's characteristics.
The use of \textit{CoreDSL} enables the integration in the retargeting compiler \textit{Seal5} and the instruction set simulator \textit{ETISS}. 
The following are contributions of the paper:


\begin{itemize}[noitemsep,topsep=0pt]
    \item The generation of new RISC-V instructions based on assembly structures.
    \item The selection of instructions according to optimization targets.
    \item The creation of a tool, interoperable with \textit{Seal5}, \textit{ETISS}, \textit{Longnail} and extendable in regard to different optimization targets and further possibilities to support different instruction widths using a RISC-V specific implementation regarding ISA specifics.
\end{itemize}



\section{State of the Art}
\label{sec:RelatedWork}




A multitude of extensions have been developed for the RISC-V architecture. 
One such extension is the general-purpose RISC-V Compressed (RVC) extension \cite{waterman2011improving}.
The objective of RVC is twofold: to minimize the size of code and to enhance energy efficiency.
It employs a one to one replacement of 32-bit with 16-bit instructions, resulting in a 25 \% reduction in code size. 
Conversely, there are application-specific extensions developed for optimized network operations \cite{sodergren2021risc,Cao2018}. 
Subsequently, there are security-specific RISC-V extension like optimizations for the Advanced Encryption Standard \cite{marshall2020design}.
\Name{} is designed to generate arbitrary application-specific extensions. 
It automates the generation process, thereby reducing the effort required to extend the ISA for a specific program. 

Atasu et al. have proposed an algorithm for identifying new instructions based on a Data Flow Graph (DFG) \cite{ConvexSubgraphEnumeration}. 
In this algorithm, constrained maximum convex subgraphs are selected and interpreted as new instructions.
Achieving a performance improvement of one order of magnitude for most of the benchmarks tested. 
Then, Galuzzi et al. have proposed linear complexity algorithms for the generation of Multiple Input Single Output (MISO) \cite{LinearComplexityMISO} and Multiple Input Multiple Output \cite{LinearComplexityMIMO}.
These algorithms are based on MAXMISO, i.e. the maximum overlap free subgraphs with a single output of the program Directed Acyclic Graph (DAG) representing the new MISO instructions \cite{MAXMISO}.
In addition, Galuzzi et al. have also proposed an algorithm for the automated instruction selection of application-specific ISEs \cite{ISESel}.
The instruction selection is formulated as an Integer Linear Programming (ILP) problem with the objective of minimizing the execution time and the constraint of the hardware area.
Their approach has yielded an acceleration of up to x3.7.
Our approach aligns with the majority of prior works on the generation of MISO instructions. 
However, the generation process in this case is distinct in that it is based on parsed assembly code, leveraging its structures.

There are proprietary and open source retargetable compilers.
The proprietary compilers include \textit{Codasip Studio} developed by \textit{Codasip} \cite{CodasipComp}.
It is based on the LLVM pipeline.
New ISEs must be described in the company's own and application-specific high-level description language \textit{CodAL}.
Conversely, there are also open source tools that provide automated support for ISE.
Hepola et al. proposed \textit{OpenASIP} in their work \cite{openASIP2}.
This toolset is designed to primarily generate RTL for custom instructions.
Additionally, it incorporates a retargetable compiler that allows to compile custom instructions.
Finally, van Kempen et al. proposed the retargeting compiler \textit{Seal5} \cite{Seal5}.
This tool generates global ISel patterns in addition to the aforementioned and uses the open \textit{CoreDSL} description language as an interface.
In addition, this tool allows the use of immediate values with variable bit length and generates tablegen structures \cite{Seal5VarImm}.


\section{Automated Instruction Set Extension}
\label{sec:AISE}

\begin{figure}[ht!]
    \begin{center}
        \includegraphics[width=\textwidth]{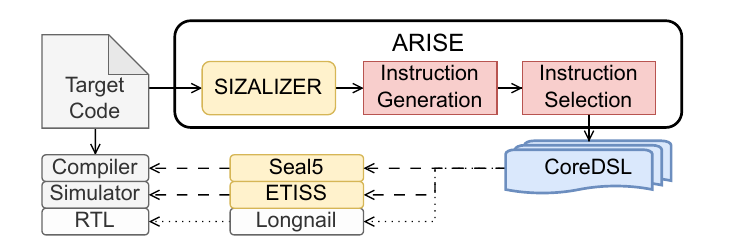}
        \caption{Tool Architecture.} \label{fig:Architecture}
    \end{center}
\end{figure}

\Name{} is a tool for automated ISE generation.
It uses \textit{SIZALIZER} to parse and preprocess target assembler code \cite{SIZALIZER}.
In addition, the code is initially executed and written to a trace file, which is parsed in and set in a one-to-many connection with the previously parsed assembly objects.
All used standalone tools are shown in yellow boxes in Figure \ref{fig:Architecture}.
The red boxes represent the contributions of \Name{}.
The code structures provided by \textit{SIZALIZER} are passed to the \textit{Instruction Generation}. 
In this step, the search is conducted for structures within the existing code that can be encoded in 32 bits.
The generated instructions are then passed to \textit{Instruction Selection}.
They are evaluated using the proposed metrics.
The highest ranking, are then selected and written to output files in CoreDSL format.

\subsection{Instruction Generation}

Instruction generation is a function that, given a list of instructions, deterministically outputs a set of new instruction proposals.
To work around this, \Name{} uses a greedy algorithm to generate maximal instructions with the restriction to MISO, similar to MAXMISO \cite{MAXMISO}.
Unlike MAXMISO, \Name{} is based on instruction sequence selection at assembly level.
This allows it to be close to the ISA because the abstractions are already resolved.
Initially, it generates a new and empty instruction.
The empty instruction is then extended by a sequence of instructions having a data flow connection.
The new instructions are marked as closed as soon as the amount of freely encodeable bits is exhausted.
In addition, a new instruction can be marked as closed if an output is consumed elsewhere, for example by saving an output operator or overwriting an input operator.
Furthermore, control flow instructions terminate the new instructions.
The set of all completed instructions now comprises all generated instructions.
These are then post processed.
Duplicates and replaceable ones are removed.
For example, if an instruction is generated with an immediate value of fewer bits than an equal instruction with more bits, the former is removed.
However, there is no logic solver that would allow resolving logical equivalences.
Thus, there is no complete removal of all isomorphic instructions.

\lstdefinestyle{customasm}{
    basicstyle=\ttfamily\footnotesize,
    numbers=left,framexleftmargin=5mm,
    numberstyle=\color{black},
    stepnumber=1,
    numbersep=5pt,
    backgroundcolor=\color{white},
    frame=topbottom,
    breaklines=true,
    escapeinside={(*@}{@*)},
    morekeywords={mov, add, sub},
    xleftmargin=15pt
}

\lstset{style=customasm}

%

\begin{lstlisting}[caption={Instruction Generation Example},captionpos=b,label={lst:InstructionGeneration}]
    (*@\color{blue}add	    t1, s0, a5@*)
    (*@\color{blue}add	    a5, t1, a2@*)
    (*@\color{teal}xori	a1, a5, 0x1@*)
    (*@\color{teal}c.or	a0, a1@*)
    (*@\color{teal}sltu	s3, zero, a0@*)
    addi	a0, a0, 0x80
\end{lstlisting}

The sample assembly code in Listing \ref{lst:InstructionGeneration} shows how the instruction generation would merge a given code into new instruction units.
The generated instructions are indicated by the different colors. 
These instructions are then generalized and, immediate widths are derived.
For example, the last union, colored in teal, becomes a \textit{xori\_or\_sltu rd, rs1, rs2, rs3, imm[2:0]}, where 9 bits are used for the opcode, 20 bits for the source and destination registers, and 3 bits for the immediate.
The \textit{addi} is not additionally encodeable in this instruction.


\subsection{Instruction Selection}

The term \textit{Instruction Selection} refers to the selection of a subset of the generated instructions.
\Name{} contains predefined metrics for static and dynamic code size optimization and the dynamic instruction count.

\begin{algorithm}[t]
    \caption{Instruction Improvement Metrics Template}
    \label{algo:StaticSize}
    \begin{algorithmic}[1]
        \Require $instructions$: List[Instruction], $pattern$: Pattern, \Call{Metric}{}: Function(Pattern$\mid$List[Instruction]) $\rightarrow$ Integer
        \Ensure $improve$: Integer

        \State $improve, index \gets 0$

        \While{$index <= \Call{size}{instructions} + \Call{size}{pattern}$}
            \State $instr\_slice \gets instructions[index:index+\Call{size}{pattern}]$
            \If{\Call{MatchPattern}{instr\_slice, pattern}}
                \State $improve \gets improve + \Call{Metric}{instr\_slice} - \Call{Metric}{pattern}$
                \State $index \gets index + \Call{size}{pattern}$
            \Else
                \State $index \gets index + 1$
            \EndIf
        \EndWhile
        \State \Return $improve$
    \end{algorithmic}
\end{algorithm}

The metrics are implemented in the instruction improvement metrics template as depicted in Algorithm \ref{algo:StaticSize}.
For each generated instruction, a function is passed that calculates a value based on the goal.
It calculates the code size in bytes, and, for the count reduction, it counts the instructions.
This is executed for all given instructions, and the delta is added up.
The Algorithm \ref{algo:StaticSize} then returns the weight.
This metric is then called for each pattern and evaluated.
The number of instructions to be selected is calculated from the opcode length as follows: $sel\_count \gets 2^{opcode\_bits - custom\_bits} * custom\_count$.
The count of custom opcode ranges is equal to 4, and their opcode bit count is equal to 7.
The opcode bits are be greater than or equal to the base opcode bits. 
The length of the opcode and therefore the number of instructions to be generated in \Name{} is static.

To achieve an improved performance in dynamic cases, \textit{SIZALIZER} offers the possibility to process the instructions of the trace file in a one-to-many relation of static assembler instructions to the dynamic execution points.
The metric then follows the relation for each matched pattern to calculate the improvement.
This simplification is possible because there are no patterns that go beyond basic blocks. 
Therefore, the runtime of the selection is formed by the number of instructions in the assembly, which is usually many orders of magnitude smaller than the number of instructions in the trace. 


\section{Evaluation}
\label{sec:Evaluation}

\Name{} is evaluated using \textit{Embench-Iot} \cite{Embench}.
The benchmarks are compiled with the RISC-V standard \textit{IMACFD} extension and the highest code size compression optimization \textit{-Oz} with the retargeting compiler of \textit{Seal5} in order to be comparable with the extended variant.
Based on this, \Name{} generates up to 16 instructions with an opcode length of 9 bits for each optimization target.
These instruction set extensions are then written to \textit{CoreDSL} files.
\textit{Seal5} and \textit{EITSS} are extended using the generated extension.
The code is then executed and compared to the unmodified variant.


\begin{figure}[ht]
    \centering
    \begin{tikzpicture}

        \definecolor{darkgray176}{RGB}{176,176,176}
        \definecolor{gray}{RGB}{180,180,180}
        \definecolor{steelblue31119180}{RGB}{31,119,180}

        \begin{axis}[
            tick align=outside,
            tick pos=left,
            x grid style={darkgray176},
            ybar,
            symbolic x coords={aha-mont64,crc32,cubic,edn,huffbench,matmult-int,md5sum,minver,nbody,nettle-aes,nettle-sha256,nsichneu,picojpeg,primecount,qrduino,sglib-combined,slre,st,statemate,tarfind,ud,wikisort},
            xtick=data,
            x tick label style={rotate=25, anchor=east},
            ylabel={\% Static Size},
            y grid style={gray},
            ymin=-3.7, ymax=6.8,
            ymajorgrids,
            yminorgrids,
            bar width=5pt,
            enlarge x limits=0.03,
            nodes near coords,
            every node near coord/.append style={font=\tiny}
        ]
            \addplot coordinates {(aha-mont64, 2.4) (crc32, 0.5) (cubic, 0.0) (edn, 4.7) (huffbench, 2.2) (matmult-int, 1) (md5sum, 0.5) (minver, 2.0) (nbody, 1.1) (nettle-aes, 0.0) (nettle-sha256, 4.5) (nsichneu, 0.1) (picojpeg, -1.4) (primecount, 0.5) (qrduino, 3.7) (sglib-combined, 1.0) (slre, 0.9) (st, 0.0) (statemate, 0.1) (tarfind, 1.3) (ud, 3.6) (wikisort, 4.2)};
        \end{axis}
    \end{tikzpicture}
    \caption{Static Size Improvement}
    \label{fig:StaticSize}
\end{figure}
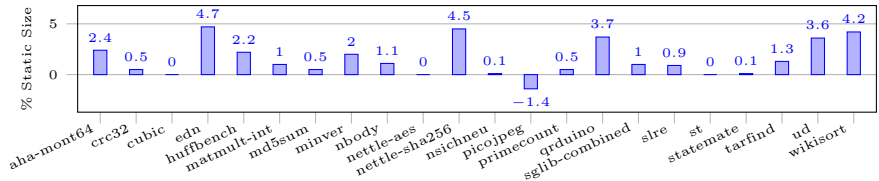

Figure \ref{fig:StaticSize} shows the static code size improvements in percent.
It is calculated based on the absolute static code size read from the unchanged and the instrumented binary file.
These sizes are adjusted by the same functions and instructions as during generation.
The improvement is then calculated on the basis of these base values.
It shows a significant improvement in static code size for the majority of benchmarks, on average 1.48 \%.
In particular, \textit{edn}, \textit{nettle-sha256} and \textit{wikisort} stand out with an improvement of over four percent.
However, for \textit{nettle-aes}, \textit{nsichneu}, \textit{st} and \textit{statemate}, none or no significant influence on the code size is measured.
The case of \textit{picojpeg} shows a negative influence on the code size.
This phenomenon can be attributed to the effects of the compiler instruction selection, which \Name{} has not taken into account.
These effects include, e.g.  more registers are captured in the instruction input for a single instruction, it can occur that more registers are reserved until the execution instruction, and therefore some previous instructions cannot be compressed.

\begin{figure}[ht]
    \centering
    \begin{tikzpicture}
            
        \definecolor{darkgray176}{RGB}{176,176,176}
        \definecolor{gray}{RGB}{180,180,180}
        \definecolor{steelblue31119180}{RGB}{31,119,180}

        \begin{axis}[
            tick align=outside,
            tick pos=left,
            x grid style={darkgray176},
            ybar,
            symbolic x coords={aha-mont64,crc32,cubic,edn,huffbench,matmult-int,md5sum,minver,nbody,nettle-aes,nettle-sha256,nsichneu,picojpeg,primecount,qrduino,sglib-combined,slre,st,statemate,tarfind,ud,wikisort},
            xtick=data,
            x tick label style={rotate=25, anchor=east},
            ylabel={\% Dynamic Size},
            y grid style={gray},
            y grid style={gray},
            ymin=-11, ymax=22,
            ymajorgrids,
            yminorgrids,
            bar width=5pt,
            enlarge x limits=0.03,
            nodes near coords,
            every node near coord/.append style={font=\tiny},
        ]
            \addplot coordinates {(aha-mont64, 13.1) (crc32, 3.3) (cubic, 0.0) (edn, 8.1) (huffbench, 2.0) (matmult-int, 16.0) (md5sum, 7.1) (minver, 2.2) (nbody, 0.3) (nettle-aes, 0.8) (nettle-sha256, 5.8) (nsichneu, 0.0) (picojpeg, -5.1) (primecount, 7.1) (qrduino, 8.) (sglib-combined, 3.8) (slre, 0.8) (st, 0.0) (statemate, 0.0) (tarfind, 2.9) (ud, 2.6) (wikisort, 0.9)};
        \end{axis}
    \end{tikzpicture}
    \caption{Dynamic Size Improvement}
    \label{fig:DynamicSize}
\end{figure}
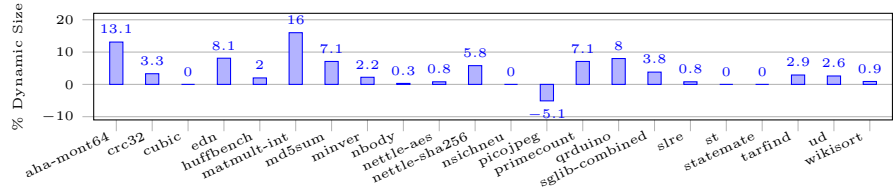

The dynamic code size improvement is shown in Figure \ref{fig:DynamicSize} in percent.
This is calculated in the same way as the static code size, except that the execution information is analyzed instead of the static binary file.
It contains many orders of magnitude more instructions, with an instruction distribution shifted around the execution focus.
This leads to an expected scaling of efficiency, which occurs when instructions are found in the blocks that are executed very frequently.
For example, the \textit{matmult-int} benchmark has a dynamic improvement of \textit{16 \%} with a static improvement of \textit{1 \%}.
This amplification effect also occurs with \textit{picojpeg} in the negative direction.
However, this can also lead to a reduction in efficiency of \Name{} if only instructions are found in less frequently executed blocks.
This is the case for \textit{ud} and \textit{wikisort}.
The average reduction in dynamic code size is therefore 3.84 \%.

\begin{figure}[ht]
    \centering
    \begin{tikzpicture}
        
        \definecolor{darkgray176}{RGB}{176,176,176}
        \definecolor{gray}{RGB}{180,180,180}
        \definecolor{steelblue31119180}{RGB}{31,119,180}
            
        \begin{axis}[
            tick align=outside,
            tick pos=left,
            x grid style={darkgray176},
            ybar,
            symbolic x coords={aha-mont64,crc32,cubic,edn,huffbench,matmult-int,md5sum,minver,nbody,nettle-aes,nettle-sha256,nsichneu,picojpeg,primecount,qrduino,sglib-combined,slre,st,statemate,tarfind,ud,wikisort},
            xtick=data,
            x tick label style={rotate=25, anchor=east},
            ylabel={\% Dynamic Count},
            y grid style={gray},
            ymin=0, ymax=30,
            ymajorgrids,
            yminorgrids,
            bar width=5pt,
            enlarge x limits=0.03,
            nodes near coords,
            every node near coord/.append style={font=\tiny}
        ]
            \addplot coordinates {(aha-mont64, 17.3) (crc32, 9.1) (cubic, 0.0) (edn, 17.6) (huffbench, 4.0) (matmult-int, 24.2) (md5sum, 13.1) (minver, 3.0) (nbody, 0.0) (nettle-aes, 16.5) (nettle-sha256, 13.5) (nsichneu, 0.0) (picojpeg, 0.4) (primecount, 13.1) (qrduino, 12.1) (sglib-combined, 5.9) (slre, 0.9) (st, 0.0) (statemate, 0.0) (tarfind, 4.3) (ud, 5.5) (wikisort, 2.1)};
        \end{axis}
    \end{tikzpicture}
    \caption{Dynamic Instruction Count Improvement}
    \label{fig:DynamicCount}
\end{figure}
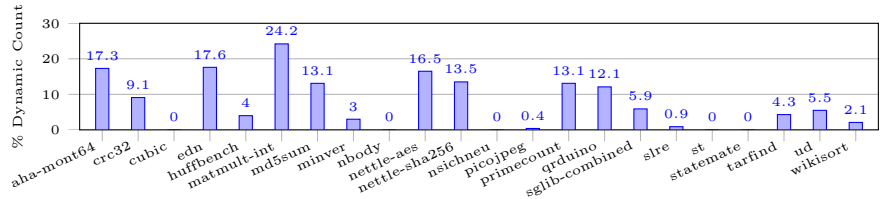

Figure \ref{fig:DynamicCount} shows the reduction in the number of instructions executed.
These are adjusted by the same factors as in the evaluation for the static and dynamic code size evaluation.
It can be seen here that there is no increase in the number of instructions, because optimizing instructions replace at least one and thus is always improving under this metric.
Apart from this, it can also be seen that the absolute values and its distribution are an amplification of those of the dynamic code size.
Particularly noteworthy is the 24.2 \% reduction in the number of instructions to be executed for \textit{matmult-int}.
However, there is also no significant improvement in some cases, which are analogous to those for the dynamic size.
The average reduction in instructions is 7.39 \%.



\section{Discussion}
\label{sec:Discussion}


With \Name{} a tool is presented that can automatically generate instructions based on given software.
It is also shown that, on average, an improvement is achieved.
However, the negative values also show that effects that cannot be taken into account by \Name{} may lead in the opposite direction. 
Therefore, it may be necessary to manually reprocess the generated instructions to achieve optimal results.
This cannot be avoided without considerable effort on the part of \Name{}.
A complete search would require considering all combinations of instructions for an optimum.
This would impose an exponential runtime. 
The efficient approach would be to implement metrics that take into account effects such as outlining, compressibility, register capturing, etc., and thus provide better or at least non-negative results.

The evaluation also shows that instructions can be generated without a significant impact on the selected goal. 
Because in the \textit{Instruction Generation}, the first possible combination of instructions is generated.
Other and possibly better instruction combinations are missed.
However, the power of the set of all possible combinations also grows exponentially in this case.
This could be limited with the restriction that each basic block is considered separately, since \Name{} is limited to arithmetic, and logical instructions anyway.
In addition, this may be done for the most relevant blocks by counting the frequency of execution.

Finally, no hardware-related metrics have not yet been considered and evaluated.
These include the cycle count, area or energy consumption.
To implement and evaluate these, it is necessary to generate RTL code for the ISE. 
\textit{Longnail}, represented in Figure \ref{fig:Architecture} with a light gray background, would be suitable for this.
It provides a high-level synthesis of ISE for RISC-V processors from \textit{CoreDSL} Descriptions \cite{Longnail}.
However, this tool is not open source.
Thus, it could not be integrated in \Name{}.

\section{Conclusion}
\label{sec:Conclusion}

In this paper, we introduced \Name{}, a tool that bridges the gap between RISC-V ISA and software for embedded systems. 
ARISE automates the generation of custom instructions based on assembly structures.
It incorporates metrics-based instruction selection, enabling precise optimization for static and dynamic code size and instruction count reduction. 
By encoding the ISE in the high-level ISA description language \textit{CoreDSL}, ARISE ensures seamless integration into advanced toolchains such as \textit{Seal5} and \textit{ETISS}. 
Experimental results show significant improvements, including average reductions in static code size by 1.48 \%, dynamic code size by 3.84 \%, and number of instructions executed by 7.39 \% for the Embench-IoT benchmarks. 
These results highlight the potential of \Name{} to enable tailored ISA optimization in embedded applications.

%
%

\bibliographystyle{splncs04}
\bibliography{main}

\end{document}